# APPLICATION AND NETWORK LAYERS DESIGN FOR WIRELESS SENSOR NETWORK TO SUPERVISE CHEMICAL ACTIVE PRODUCT WAREHOUSE

Ahmed Zouinkhi, Kais Mekki, Mohamed Naceur Abdelkrim

Research unit of Modeling, Analysis and Control of Systems (MACS)
National Engineering School of Gabes, rue Omar Ibn Elkhattab, 6029 Gabes, Tunisia.

## ABSTRACT

*Wireless sensor networks have profound effects on many application fields like security management which need an immediate and fast system reaction. Indeed, the monitoring of a dangerous product warehouse is a major issue in chemical industry field. This paper describes the design of chemical warehouse security system using the concept of active products and wireless sensor networks. A security application layer is developed to supervise and exchange messages between nodes and the control center to prevent industrial accident. Different security rules are proposed on this layer to monitor the internal state and incompatible products distance. If a critical event is detected, the application generates alert message which need a short end to end delay and low packet loss rate constraints by network layer. Thus, a QoS routing protocol is also developed in the network layer. The proposed solution is implemented in Castalia/OMNeT++ simulator. Simulation results show that the system reacts perfectly for critical event and can meet the QoS constraints of alert message.*

## KEYWORDS

*Security System, Active Product, Wireless Sensor Networks, Routing protocol, Quality of Service, Castalia Simulation*

## 1. INTRODUCTION

In recent years, the rapid technological advances in low power and highly integrated digital electronics, small scale energy supplies, and low power radio technologies have created low cost and multifunctional intelligent devices. These devices are equipped with small battery, a tiny microprocessor, a radio transceiver and a set of sensor modules that used to acquire information of their surrounding environment. These technological advances lead to different new research field such as Internet of Things [1] and Active Product [2].

The active intelligent product had large expansion on the industry. Through this concept, the industrial product is not simply a physics entity but it is an active object which is able to communicate and exchange information with the systems. These capacities were made with the RFID identification technology (Radio Frequency Identification) [3] and the Wireless Sensors Networks (WSN). However unlike RFID, WSN allows going beyond the information exchange to the sensing of ambient environment parameters (e.g. temperature, pressure, humidity…) [4].

This concept attracted the interest of several research projects. As example, COBIS project (Collaborative Business Items) [5,6] has developed a new approach for business processes involving physical entities such as goods and tools in enterprise. COBIS improved the networked systems to create cooperative products named particles, used for various applications particularly for monitoring the industrial products. TecOlab from Karlsruhe University and MIT [7] has announces the concept of active physical documents for the integrity management of written files to restrict accesses and keep the track of the document changes. It introduces the DigiClip system that provides a solution to convert passive paper documents to active physical entities. Moreover, CHAOS project [8] also employs the intelligent object approach to secure the exchange of information in distributed systems.

The security of industrial chemical products is an important issue. So, the using of active product technology may facilitate its supervising process. The security of such product involves the internal reaction against the ambient values changes (e.g. temperature) and the incompatibility of chemical substances during the storage and transport phases.

This paper proposes a security system of chemicals products warehouse. Such storage products may cause critical danger if security rules are not respected. The container of chemical substances is transformed into a communicating entity (i.e. active product) by embedding a wireless sensor node. Thus, the container could supervise its internal state and the external changes of its environment (e.g. temperature, brightness, and humidity). Moreover, the containers could control the distance between them by periodic message exchange (RSSI method [9]) to prevent the closeness of the incompatible chemical substance. If there are critical and alert events (e.g. high temperature, incompatible products are very close), the node is able to make decisions and sends alert messages to the control center (sink node).

The developed system uses a centralized approach and point-to-point connection for the control center to communicate with nodes using a specific security application layer of WSN (i.e. the sink communicates directly with the nodes in the warehouse). However, the supervised nodes use a multi-hops connection to communicate and send messages to the control center (i.e. the messages passes from one node to another until it reaches the sink). This last case requires a routing protocol development in the network layer of sensor nodes. Indeed, the security application layer generates two classes of message: *alert message* and *routine message (i.e. daily exchanged message)*. An alert message should be encapsulated in high priority packet in the network layer and routed across the WSN toward the sink node. Such packet requires strict constraints on both delay and loss ratio in order to report the message to the control center within certain time limits without loss. These performance metrics (delay and loss ratio) are usually referred to as Quality of Service (QoS) requirements [10]. Thus, enabling high priority packet in the network layer requires QoS based routing protocol. Many QoS based routing protocols specifically designed for WSN have been proposed in the literature, for example, SPEED [11], AODV [12], RPAR [13], THVR [14], RRR [15], EQSR [16], and MARP [17]. For our chemical security system, we had developed routing strategies for each classe of messages. Disjoint multi-paths routing is proposed for alert message. However, the routing algorithm employs neighborhood gathering information and next node selection phase for routine message.

These developed application and network layers have been implemented and simulated using Castalia/OMNeT++ tools. The security system is then evaluated by studying the message delivery ratio, end to end delay, and alert detection validation.

The rest of paper is organized as follows. In the next section, the communication topology of the security system is presented. In section 3, the security rules are detailed. And, section 4 and 5 present the application and network layers design, respectively. Then in section 6, the performance of the proposed solution is analyzed. Finally, section 7 concludes the paper.

## 2. COMMUNICATION TOPOLOGY OF THE ACTIVE CHEMICAL PRODUCT SECURITY SYSTEM

### 2.1. The active chemical product

The concept of active product consists in endowing a product with the capacities of communicating, informing, acquiring, deciding, and reacting to the changes of its ambient environment [18,19]. This concept is used in our security system by integrating wireless sensor node in every chemical container as shown in figure 1. Thus, each container is upgraded with sensing, computation, interaction, and communication capacities.

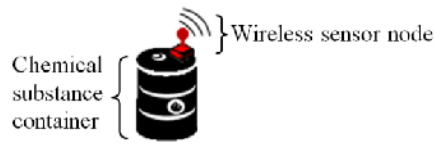

Figure 1. Transformation of the chemical container into active product

### 2.2. The network communication topology

Our active security system adopts a point to point approach for the control center to send message to the supervised nodes through the sink. The sink is a powerful node with more sophisticated resources. It has improved processing, storage, energy, and communication power. So, the sink sends messages directly to the supervised node without multipath connection as shown in figure 2 where S and D are the source node and the destination node of a message, respectively.

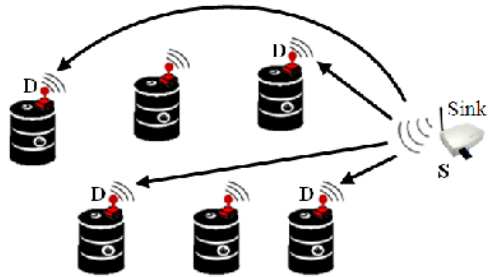

Figure 2. Communication topology for the sink

As for the container, it uses resource constrained nodes and much less expensive than the sink. Such nodes have limited energy resource (battery) which requires low transmission powers to increase its life time. Thus, it uses short transmission range for message exchange (i.e. wireless transmission consumes more energy than any other communication activity [20,21]). For this, a multi-hops connection is used for container nodes to send message to the sink. The messages passes from one node to another using routing protocol until it reaches the control center as shown in figure 3.

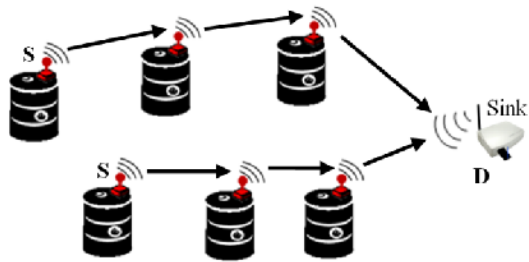

Figure 3. Communication topology for the container

## 3. SYSTEM SECURITY RULES

The system security rules define the chemical substance state of each supervised container. The variation of certain ambient environment such as the temperature could lead to critical industrial accident. Moreover, different chemical substances are stored in the same warehouse and some substances could be incompatible with others. For this reasons, the following critical chemical factors are considered:

- *The ambient temperature*: some chemical substances could lead to dangerous explosion or flame. So through temperature thresholds (min and max), the internal state of the container could be supervised.
- *The product incompatibility*: the distance between containers is supervised to prevent all possible dangerous reaction of the incompatible chemical substances.

To control these factors, the following security rules are developed: *static*, *dynamics*, *community*, and *global* rules.

### 3.1. Static rules

The container node employs temperature sensor. In order to avoid critical reactions, the static rule requires that the ambient temperature value didn't exceed a threshold values (min and max). This threshold is defined according to the chemical substance characteristics.

So, each sensed temperature value *Vsr* is characterized by two critical values $Vsr_{min}$ and $Vsr_{max}$ associated with a safety margin $\Delta Vsr$. For each node *i*, the security level $S_{sr}$ of the static rules is evaluated as follow:

- *G*: if the value of the sensor *i* defines a good state.
- *B*: if the value announces a bad state.
- *D*: if the value indicates a dangerous state.

The statics rules $S_{sr}$ are presented by the equation 1.

$$S_{sr} = \begin{cases} G & if \quad Vsr \in \;]Vsr_{min} + \Delta Vsr \,, Vsr_{max} - \Delta Vsr\;[ \\ B & if \quad Vsr \in [Vsr_{min}, Vsr_{min} + \Delta Vsr\,] \cup [Vsr_{max} - \Delta Vsr, Vsr_{max}] \\ D & if \quad Vsr \in \;]-\infty, Vsr_{min}[ \cup\;]Vsr_{max}, +\infty[ \end{cases} \quad (1)$$

International Journal of Computer Science, Engineering and Applications (IJCSEA) Vol.4, No.6, December 2014

## 3.2. Dynamic rules

The purpose of these rules is to supervise the temporal variation of the static rules. If a bad state persists for period of time, a dangerous state is announced. Moreover, a counter is implemented to gather the switch between good and bad states. If the counter exceeds a threshold value *nc*, a danger state is announced. The dynamic rules $S_{dr}$ are presented by the equation 2.

$$S_{dr} = D \text{ if } \begin{cases} \exists t_1, \forall t \in [t_1, t_1 + T_{cr}] \; S_{sr}(t) = B \\ or \\ Occur(S_{sr}(t)) \geq n_c \end{cases} \quad (2)$$

With:

- $T_{cr}$: is a critical period of time. It is fixed according to the chemical product that not overcomes when a bad state is reached.
- *nc*: is the number of authorized $S_{sr}$ switch between states *G* and *B*.
- *Occur(x)*: is a function initialized to zero and incremented when $S_{sr}$ switch from state *G* to *B*.

## 3.3. Community rules

Some products could have incompatibility constraints according to their chemical characteristics. Such constraint is proportional to the distance between chemical products (container). For example, the chemical symbols *H2SO4* and *HF* are incompatible and represent a danger if the distance between them is under a predefined threshold value [19]. The community rules $S_{cr}$ could lead to one of the three states as described in equations 3, 4, and 5.

$$S_{cr} = G \text{ if } \begin{cases} \begin{cases} F_{comp}(Symb_i, Symb_j) = Incompatible \\ and \\ D(P_i, Pj) > D_{min} + \Delta D \end{cases} \\ or \\ F_{comp}(Symb_i, Symb_j) = Compatible \end{cases} \quad (3)$$

$$S_{cr} = B \text{ if } \begin{cases} F_{comp}(Symb_i, Symb_j) = Incompatible \\ and \\ D(P_i, Pj) \subset [D_{min}, D_{min} + \Delta D] \end{cases} \quad (4)$$

$$S_{cr} = D \text{ if } \begin{cases} F_{comp}(Symb_i, Symb_j) = Incompatible \\ and \\ D(P_i, Pj) < D_{min} \end{cases} \quad (5)$$

With:

- $F_{comp}(x,y)$: is a function which studies the compatibility between the chemical symbols of two products $P_i$ and $P_j$.
- $D(x, y)$: is the distance between two products $P_i$ and $P_j$.
- $D_{min}$: is a critical distance (threshold) that must be respected between incompatible products.
- $\Delta D$: is a distance margin fixed according to the chemical characteristic of the product.

57

## 3.4. Global rules

After calculate the state of each rule (static, dynamic, and community), it is necessary to get the global security state $S_{gr}$. $S_{gr}$ describes the global state of the product as described in equations 6.
$S_{gr}= f(S_{sr}, S_{dr}, S_{cr})$

$$S_{gr} = \begin{cases} G & if \quad \forall \alpha \begin{cases} S_\alpha \neq D \\ S_\alpha \neq B \end{cases} \\ B & if \quad \forall \alpha \begin{cases} S_\alpha \neq D \\ \exists \alpha, S_\alpha = B \end{cases} \\ D & if \quad \exists \alpha, S_\alpha = D \end{cases} \quad (6)$$

With: $\alpha =\{sr, dr, cr\}$

## 4. APPLICATION LAYER PROTOCOL

This section describes the active product behavior. And, it presents the control_center/product and product/product messages exchange to achieve permanent supervising of the warehouse. The chemical product is active using the following parts: *registration*, *configuration*, *internal surveillance,* and *alert announcement*.

### 4.1. Registration part

This part is developed to register the new products in the security system. The product announces them self in the warehouse by sending a discovery message to the control center (the sink node) as shown in figure 4. To achieve this action, we used two types of messages:

- *CTR*: it announces the product in the network, it is sent continuously until the sink responds.
- *ACKCTR*: acknowledgement of the *CTR* reception by the sink.

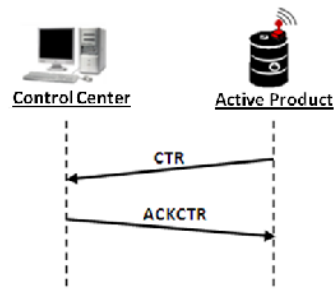

Figure 4. Sequence diagram of the registration part

### 4.2. Configuration part

Each product communicates and interacts within its neighborhood using the different security information. The active product gets this information through the configuration part by sending specific messages to the control center. In fact, the necessary information depends on the initial state of the product. The product can have a configuration already preinstalled in its memory. It can contain information about the symbols of the chemical substance, security rules, or both. According to this configuration, we used three types of messages *NCF* (No ConFiguration):

58

International Journal of Computer Science, Engineering and Applications (IJCSEA) Vol.4, No.6, December 2014

- *NCF0*: if the product has neither the security symbol nor its security rules.
- *NCF1*: if the product has only the security symbols.
- *NCF2*: if the product has only the configuration of its security rules.

These previous messages are sent continuously until the sink responds as shown in figure 5. In the other side, the control center responds with the following messages:

- *CMD1*: it contains the configuration of the product security symbols.
- *CMD3*: it contains the configuration of the different security rules.
-

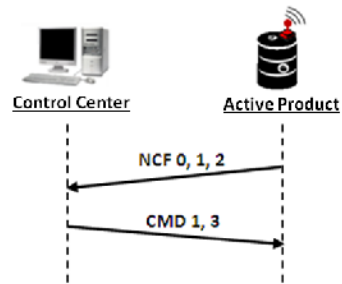

Figure 5.  Sequence diagram of the configuration part

### 4.3. Internal supervising part

Once the product is correctly configured, it becomes able to supervise its internal state and neighborhood. Any change of its environment or approximation of incompatible neighbor product must be detected using the safety rules. This supervising process is developed using the following messages:

- *GRE*: it is a greeting message periodically exchanged between neighbor products in the warehouse. It carries the product information (name, security symbols) and its current security level. It is used also to calculate the distance between two active products.
- *RSI*: it contains the received signal strength indicator (RSSI). This information is used to estimate the distance between neighbor products, and calculate the community rules. In fact, we used an equivalence-based solution between the received RSSI and the real distance value that separating two products. This equivalence is presented in figure 6. When the active product receives *RSI* message, the node calculates the RSSI value. Then, it determines the equivalent distance and verifies if this distance exceeded the threshold $D_{min}$.

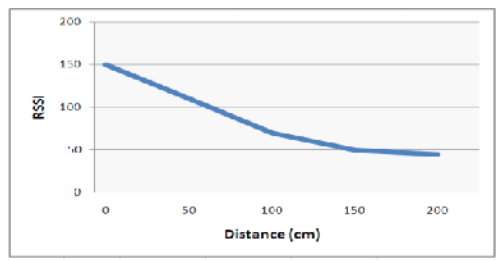

Figure 6.  Distance between products versus RSSI values

International Journal of Computer Science, Engineering and Applications (IJCSEA) Vol.4, No.6, December 2014

To achieve the internal supervising of chemical product, we used the following messages as shown in figure 7:

- *CMD2*: it is a request of the control center which requires the configuration of an active product.
- *CMD4*: it is a request of the control center to get the security rules of an active product.
- *CMD5*: it is a request of the control center to get specific ambient information of an active product.
- *CFG*: it contains the configuration of the product. It is sent after *CMD2* reception.
- *SER*: it contains the product security rules. It is sent after *CMD4* reception.
- *INA*: it carries the ambient sensed values of the product. It is sent after *CMD5* reception.

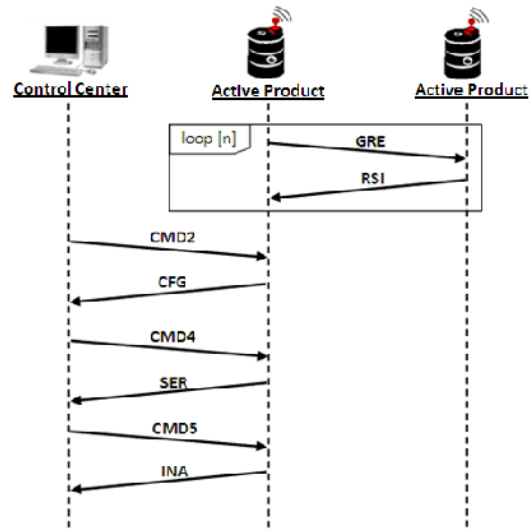

Figure 7. Sequence diagram of the internal supervising part

**4.4. Alert announcement part**

This part employs an alert message named *ALE* (ALErt) which announces a dangerous state to the control center and requires an immediate intervention as shown in figure 8.

- *ALE*: it announces the security alert to the control center. It contains the measurements that caused the alert. It is sent when the calculated security rules is considered dangerous (D) or Bad (B).
- *ACKALE*: acknowledgement of the *ALE* reception by the control center.
-

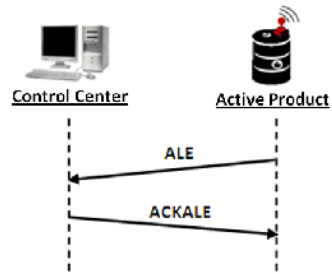

Figure 8. Sequence diagram of the alert announcement part

## 5. NETWORK LAYER PROTOCOL

*GRE* and *RSI* are broadcast messages (i.e. neighborhood exchanged messages) which don't need routing function toward the sink (control center). However, the product_node/sink communication needs routing protocol to deliver message and allow communication between them. In the following, the routing protocol is presented.

### 5.1. Gradient setup phase

The network layer protocol employs a gradient setup phase to construct the necessary information for routing path selection: *neighborhood table* and *hop count*. This phase is executed periodically since the control center begins supervising the warehouse. It works as follow: the sink broadcasts a *HELLO* packet only to its neighbors using *low transmission range* (the same transmission range of the supervised nodes), the *HELLO* packet contains: the hop count *HC* (initial value = 0) and the sender address *SA*. *HC* is used to setup the gradient to the sink (i.e. *HC* is the distance between a supervised node and the sink), and *SA* is used to build the neighborhood table of each node. After broadcasting *HELLO* by the sink, all its neighbors nodes receive this packet and each one executes the following steps:

- *If the node doesn't have a gradient:*

    1) Saves *SA* and *HC* of the sender node in its neighbors table.
    2) Increments *HC* by 1.
    3) Registers *HC* in its memory.
    4) The *HC* field is then replaced in the *HELLO* packet with the new one.
    5) Broadcasts *HELLO* to farther nodes (neighbors nodes).

- *If the node has a gradient:*

    1) Gets *SA* and *HC* of the sender node. If *SA* doesn't exist in its neighbors table, *SA* and *HC* information are saved to the table. Else, the corresponding *HC* is replaced with the new one.
    2) Increments *HC* by 1.
    3) Compares its gradient to *HC*. If the latter is smaller, the node replaces its gradient with *HC,* puts *HC* in the *HELLO* packet and broadcasts it. However, if its gradient is smaller than or equal to the HC, the node discards the packet. As a result, the gradient will memorize the short path toward the sink.

This process will continue until all the supervised nodes receive the *HELLO* packet. At that time, the gradient setup phase will be completed. As result, each node knows its distance *HC* (number of hops) to the sink, its entire neighbours and their gradient (neighbors table). Then, the nodes can starts routing packets to the control center (sink node) through the routing phase.

### 5.2. Routing phase

Our security application layer has two types of product/control_center exchanged messages: alert message (*ALE*) which requires high quality of service (e.g. short delay, low packet loss, fault tolerance), and routine messages (*CTR*, *NCF*, *CMD*,…). Thus, we had to use different strategy for the routing of such information. Moreover, in industrial environments, several faults could lead to failures of nodes in WSN. So, our routing solution has to be fault-tolerant to resist to failed nodes. The routing protocol of our security system has to take in consideration all this constraint, keeps a

International Journal of Computer Science, Engineering and Applications (IJCSEA) Vol.4, No.6, December 2014

low communication overhead, and reduces the energy consumption as much as possible. In the following, the different strategies for routing alert message and routine message are presented.

### 5.2.1. Routing of alert message

For *ALE* message, the multi-paths routing is chosen to enhance the reliability and the fault tolerance of our security system. The *ALE* source node starts multi-paths routing and creates a set of neighbors that able to forward the message towards the sink. The constructed multi-paths are node-disjoint (i.e. the multi-paths have no common nodes except the *ALE* source and the sink). Node-disjoint are usually preferred in multi-paths routing because they utilize the most available WSN resources, and hence are the most fault-tolerant [16]. If an intermediate node in a set of node-disjoint paths fails, only the path containing that node is affected. In the following, our multi-paths construction is detailed:

- *Primary path routing:* Through the neighbor table, the source selects the node which has the lowest gradient *HC* to the sink as next hop, and sends the *ALE* packet to this selected node. Similarly through its neighbor table, the next hop node of the source choose the closest node as next hop in the direction of the sink and sends out the *ALE* packet. The operation continues until sink node receives the packet as shown in figure 9.
- *Alternative path routing:* For the second alternate path, the source node sends *ALE* packet to its next most preferred neighbor (the second closest neighbor to the sink). To avoid having paths with shared nodes, each node is limited to accept only one *ALE* packet. For those nodes that receive more than one *ALE* packet, only accept the first one and reject the others. In the example of figure 9, node 9 computes it's next preferred neighbor and finds it node 7. Node 9 forwards the packet to node 7, but node 7 has been included in the primary path, then node 7 simply responds to node 9 with an INUSE packet [16] indicating that node 7 is already selected in a routing path. Immediately node 9 searches its neighboring table and selects the next preferred neighbor which will be node 5, and sends out the packet to it. Node 5 accepts the packet and continues the procedure in the direction of the sink.
-

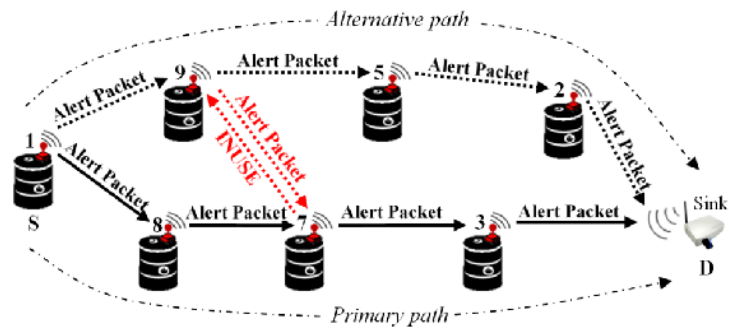

Figure 9. Node-disjoint paths routing

### 5.2.2. Routing of routine message

For routine message (CTR, NCF, CMD…), another routing strategy is developed which allows also a continuous neighbor table updating. The protocol consists of two steps for sending the message to the sink, first information gathering and then routing.

- *Information gathering:* When node needs to send routine packet to the sink, it broadcasts a request to acquire the information from the neighboring nodes in its transmission range. The

International Journal of Computer Science, Engineering and Applications (IJCSEA) Vol.4, No.6, December 2014

nodes that received the request packet send their information to the source, the response contains the residual energy of the neighboring node. Indeed, this step is used to update the neighbor table of nodes which are involved in the route toward the sink. The source node gets the *ID* of all current neighboring nodes from the received response packets, and compare between the old neighbor table and the new one. The node from the old table which is absent in the new one, are removed and the new detected nodes are added to the table.

- *Routing:* After the information gathering finishes, the routing starts. The source chooses from its neighbor table two nodes that have the lowest gradient *HC* towards the sink as shown in figure 10(a). Then, the source compare between these two nodes according to the residual energy, the nodes that had the highest residual energy level will be the next hop as shown in figure 10(b). If two nodes have the same residual energy level, the nodes that have a lower *HC* to the sink will be taken. After that, the source node sends the packet to the selected node. Upon receiving the packet, the node as next hop repeats the information gathering and routing steps to transmit the packet to another node as shown in figure 10(c). The process continues until the packet reaches the sink (control center).

-

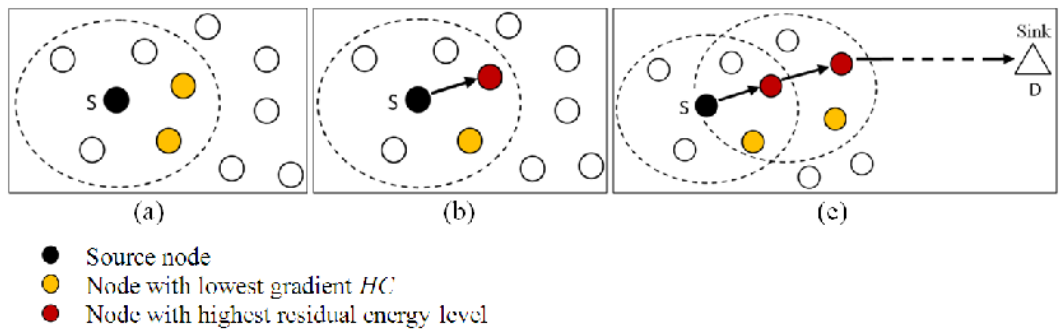

● Source node
○ Node with lowest gradient *HC*
● Node with highest residual energy level

Figure 10. Routing of routine packet

## 6. SIMULATION AND PERFORMANCE EVALUATION

In the following, the performance of the proposed security system is evaluated and discussed. The simulation settings are firstly detailed. Second, the performance results of the network layer protocol are presented and discussed. Then, different cases of application layer message exchange are studied.

### 6.1. Simulation setup

The security application and network layers were implemented with Castalia/OMNeT++ Tools. Currently, many wireless sensor network simulators are available as NS2 and SENSE but Castalia provides realistic wireless channel and radio models, and realistic node behavior especially relating to access of the radio [22].

The Crossbow's TelosB sensor node [23] is simulated which is one of the most used nodes in the industry and research. Table 1 and figure 11 show the characteristic of TelosB node.

Table 1. TelosB node characteristics

| | | |
|---|---|---|
| **Chipcon wireless transceiver** | Data rate | 250 kbps |
| | Frequency bands | 2.4 to 2.4835 GHZ |
| | Protocol | IEEE 802.15.4/ZigBee |
| **Microprocessor** | MSP430 | |
| **Flash** | 48 Kb | |
| **RAM** | 10 Kb | |
| **EEPROM** | 1 Mb | |
| **ADC** | 12 bit | |
| **Serial Com** | UART | |
| **Current draw** | Active | 1.8 mA |
| | Sleep | 5.1 µA |
| **RF power** | -24 dBm to 0 dBm | |
| **Outdoor Range** | 75 m to 100 m | |
| **Indoor Range** | 20 m to 30 m | |
| **Battery** | 2 x AA batteries | |
| **Size** | 65 x 31 x 6 mm | |
| **Weight** | 23 Grams | |

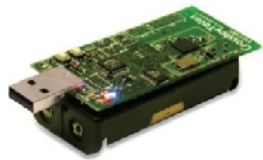

Figure 11. TelosB node

The node positions of the simulated networks (warehouse) are all uniformly distributed within a 300mx300m square (m=meters). All the nodes are stationary and the sink is located at the bottom of the square as shown in figure 12. IEEE 802.15.4 [24], a contention-based medium access control, is used as MAC protocol. The wireless radio channel characteristics such as the signal noise, interference ratio, and average path loss, are chosen to simulate the realistic modeled radio wireless channel in Castalia based on the lognormal shadowing and the additive interference models. The parameters used in this simulation study are summarized in table 2.

Table 2. Simulation parameters

| | | |
|---|---|---|
| **Warehouse area** | 300m x 300m | |
| **Nodes distribution** | Uniform random | |
| **Location of Sink (control center)** | Bottom of the warehouse | |
| **MAC layer** | IEEE 802.15.4 | |
| **Radio** | CC2420 | |
| **Sigma channel parameter** | 4 | Real radio wireless channel in Castalia |
| **Bidirectional Sigma channel parameter** | 1 | |
| **Radio Collision Model** | 2 (additive interference model) | |
| **Initial energy** | 18720 J | |
| **Simulation time** | 1000 s | |
| **Number of trials** | 60 | |

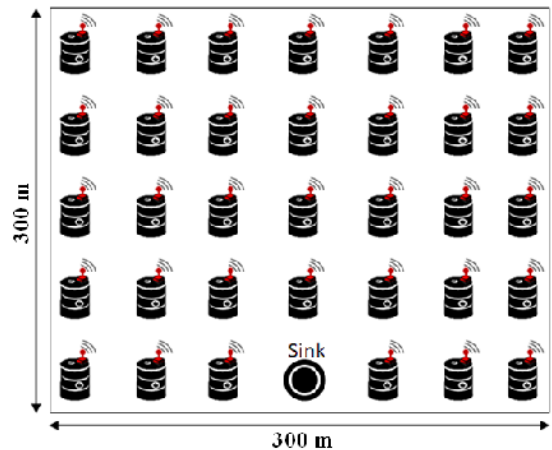

Figure 12. Nodes (active products) deployment

## 6.2. Studying the network layer performances

To study the routing protocol performances, we simulate the network under two state of traffic:

- *Not congested traffic*: the routine data packet rate is 0.2 pkt/s for each node, while the alert traffic rate is 1 pkt/s at 2 nodes.
- *Congested traffic*: the routine data packet rate is 1 pkt/s for each node, while the alert traffic rate is 5 pkt/s at 4 nodes. The network load becomes higher now as there are more sources of alert packet with higher rate.

To evaluate the performance of our routing solution, we compared its performance with Random Re-Routing protocol (RRR) [15,25] which uses also different strategy for routing of alert data and routine data. High priority packets of unusual events are routed along the preferred path (i.e. the shortest path), while the routine data packets are randomly shunted to slower and possibly longer secondary paths. In RRR, the sensor nodes change their routing policy adaptively according to the current traffic level. When the overall total traffic level is low, the preferred path will be shared by all packets. However, when the total traffic exceeds a given threshold (3 pkt/s of alert packets), the preferred path will be reserved for forwarding only the alert packets and secondary paths will be used for the routine data packets. This mechanism provides significantly better QoS (i.e. low delay and high packet delivery ratio) to alert data.

In the following, the developed QoS-based routing protocol and RRR are compared in term of: *average end to end delay* and the *packet loss ratio*.

### 6.2.1. Average end to end delay

End to end delay is an important metric in evaluating the routing protocols [10]. Figures 13, 14, 15, and 16 show the average end to end delay for each protocol and each type of packet for four levels of network density (200, 400, 600, and 800 nodes).

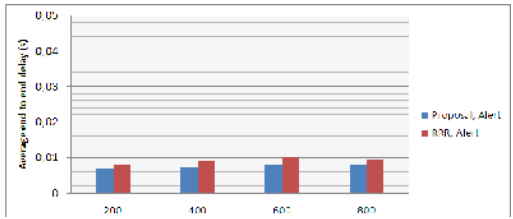

Figure 13. End to end delay in congested traffic, alert packet

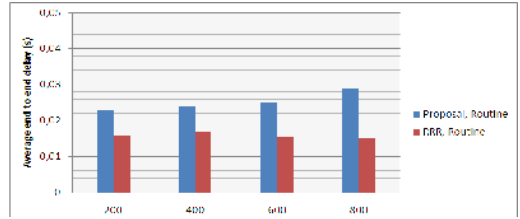

Figure 14. End to end delay in congested traffic, routine packet

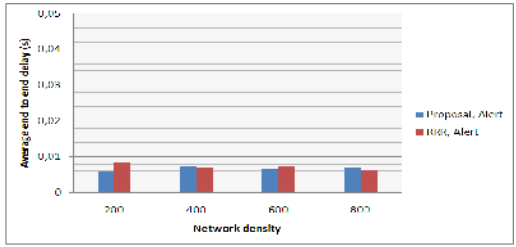

Figure 15. End to end delay in not congested traffic, alert packet

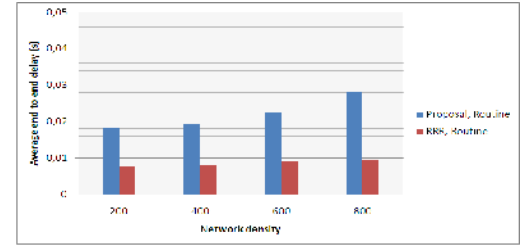

Figure 16. End to end delay in not congested traffic, routine packet

Before analyzing the performance of our routing solution, we present the result of RRR protocol. In congested traffic, the alert traffic rate at the intermediate nodes exceeds the threshold of RRR and then the alert packets are routed along the shortest path whereas the routine data packets are forwarded via random alternative paths. So as shown in figures 13 and 14, the alert data achieves lower packet delay than the routine data. Figures 15 and 16 show that the delay made for both types of data for RRR are close, because this protocol does not differentiate between types of packets in not congested traffic (similar treatment for both packets).

For our network layer, the developed protocol successfully differentiates routing services as RRR protocol. It is able to meet the time constraints for alert packets in both congested and not congested traffic as shown in figures 13 and 15, the alert packet delay is the lowest even with the increasing of the network density. RRR outperforms our protocol in the case of routing of routine data packet as presented in figures 14 and 16. Furthermore, the average delay of routine data is higher when density increases for our protocol because it suffers from a lost time on information gathering phase. In this phase and for each received routine packet, the protocol broadcasts a request and waits for responses from all the neighboring nodes to determine the next hop. So, when the number of neighboring nodes increases, the source node have to wait more time which causes more queuing and treatment delay for routine data packets despite it always chooses the paths that have the minimum number of hops to the sink. Despite this, the information gathering phase is necessary for our network layer to enhance security and fault tolerance.

In conclusion, our network layer performs better than RRR protocol in the end to end delay of alert packets but not for routine data packets.

**6.2.2. Packet Loss Ratio**

Another important metric in evaluating routing protocols is the packet loss ratio [10]. Figures 17, 18, 19, and 20 show the packet loss ratio for each protocol and each type of packet for four levels of network density (200, 400, 600, and 800 nodes).

International Journal of Computer Science, Engineering and Applications (IJCSEA) Vol.4, No.6, December 2014

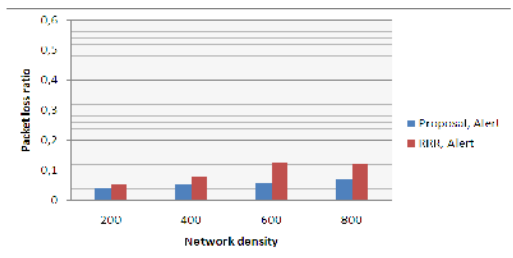

Figure 17. Loss ratio in congested traffic, alert packet

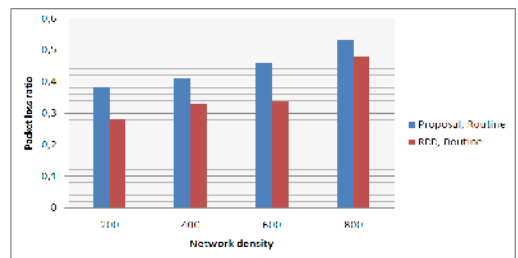

Figure 18. Loss ratio in congested traffic, routine packet

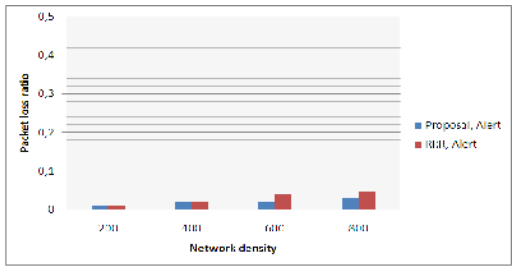

Figure 19. Loss ratio in not congested traffic, alert packet

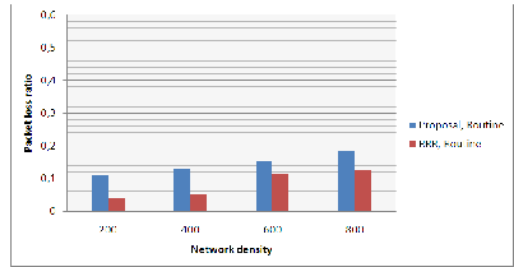

Figure 20. Loss ratio in not congested traffic, routine packet

As expected, figures 17 and 19 show that our network layer guarantees a very low packet loss rate for the *ALE* message in the two classes of traffic due to multipath routing. Discovering and maintaining multiple paths between the source and the sink node improves the routing performance by providing alternative paths. Primary and alternative paths are used simultaneously for data routing by sending multiple copies of alert packet across each path. Simultaneous multipath routing improves reliability. As long as, one of the multiple paths does not fail, the sink node will receive the alert packet.

RRR outperforms our protocol in the case of routine data packet because of the information gathering phase as delay performance. In not congested traffic, the routine data have an average packet loss rate as shown in figure 20. Lost packets increase in the case of congested traffic as shown in figure 18, the analysis of trace files of Castalia simulator showed that the majority of packets are lost due to:

- Overloading of nodes causing saturation of the queues.
- Interference because we have many transfers, so there is more concurrent access to the radio channel.

Reasoning according to the density of the network, figures 17 and 19 show that our network layer was able to maintain a low packet loss rate for alert packets due to multipath routing. For routine data, the number of lost packets increases because this protocol uses broadcast in the information gathering phase which causes many interference in case of high density.

In conclusion, our network layer is more efficient than RRR protocol because it ensures a very low packet loss rate for alert packets in both congested and not congested traffic but not for routine data packets.

## 6.3. Studying cases of application layer message exchange

### 6.3.1. Registration and configuration scenarios

Each node should be configured before starting supervising its internal state and environment. First, node sends the *CTR* message to the control center and waits to be acquitted. After receiving the *ACKCTR* message, it requests the control center to be configured by sending the *NCF* message and waits until it receives the *CMD1* or the *CMD3* messages (it depends on the type of *NCF* message *NFC0*, *NFC1*, or *NFC2*). Then, the node is considered as configured and begins supervising its internal state and its environment.

Figure 21 shows that node1 exchanges message with node0 (the sink) to be configured. Node1 started at 0.050996s when it has received the *APP_NODE_STARTUP* message of Castalia simulator, then it is configured at 0.414901s after receiving the *NCF2* message from the control center.

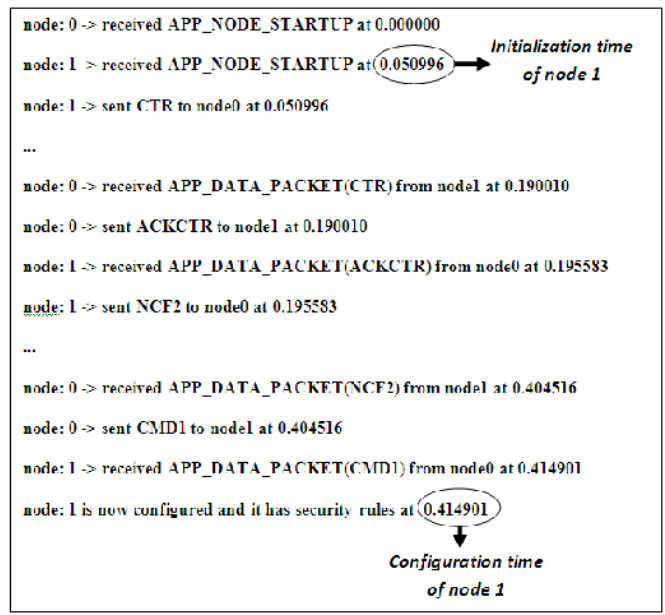

Figure 21. Castalia trace file of registration and configuration scenarios

### 6.3.2. Supervising scenarios

After registration and configuration, the node is able to supervise and react to possible environment and internal chemical changes. And, it starts periodically broadcasting the *GRE* message within its neighborhood. In the following, the system behavior and messages exchange in case of alert are presented.

- *Supervising scenario of high temperature alert:*

Figure 22 shows the simulation scenario when the internal temperature of some nodes exceeds a threshold called *VMax*. *VMax* is fixed to 14 and the node3 is configured in order to exceed this value.

International Journal of Computer Science, Engineering and Applications (IJCSEA) Vol.4, No.6, December 2014

The sensed temperature *Value* of node3 is initialized to 7 and then increased by 2 every *APP_SAMPLE_INTERVAL* time (i.e. the sampling period time of Castalia simulator). At 41.637175s, *Value* reaches the threshold *VMax*, and then node3 sent *ALE* message to node0 (the sink). The *ALE* reception is acquitted by the sink after 0.050915s.

```
node: 3 Value = 7.038390  VMax = 14.000000
...
node: 3 Value = 7.840240  VMax = 14.000000
...
node: 3 Value = 8.806861  VMax = 14.000000
...
node: 3 Value = 10.787591  VMax = 14.000000
...
node: 3 Value = 11.902659  VMax = 14.000000
...
node: 3 Value = 12.927368  VMax = 14.000000
...
node: 3 Value = 15.015800  VMax = 14.000000
node: 3 -> sent ALE Value = 7.038390 to node 0 at 41.637175
node: 0 -> received APP_DATA_PACKET(ALE) from node3 at 41.642045
node: 0 -> sent ACKALE to node3 at 41.642045
node: 3 -> received APP_DATA_PACKET(ACKALE) from node0 at 41.688090
```

Figure 22. Castalia trace file of high temperature alert scenario

- *Scenario of incompatible products alert:*

In this section, the simulation scenario of the incompatible chemical products in the warehouse is studied. For this purpose, node3 and node6 are configured to be incompatible where node3 is static and node6 is mobile. Three cases may appear:

- Case 1: the distance between the two products is safe as shown in figure 23 (i.e. node3 is far from node6). In this case, the network does not generate an *ALE* message.

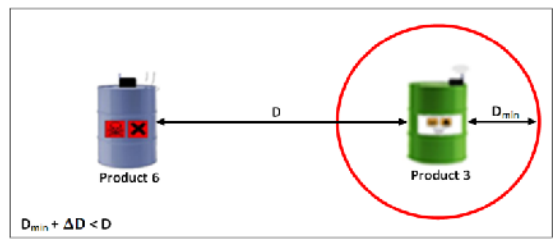

Figure 23. Safe distance between incompatible products

- Case 2: the distance between the two products is bad as shown in figure 24. Figure 24 shows that product3 (node3) is bound by two circles. The red circle presents its danger zone, where the blue one presents its bad zone. In this case, the product6 is in the bad zone (*B* state) of the product3, so an *ALE* message will be sent to the control center.

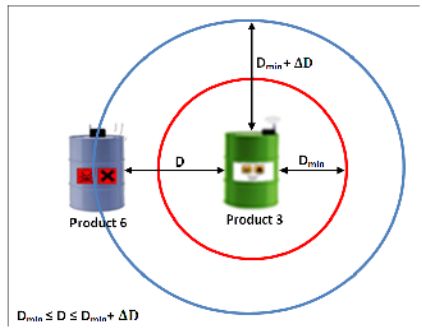

Figure 24. Bad distance between incompatible products

- Case 3: the distance between the two products exceeds the threshold $D_{min}$ as shown in figure 25.

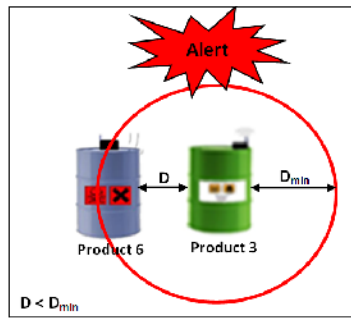

Figure 25. Dangerous distance between incompatible products

Product3 and product6 are very close (the red circle is the dangerous zone of the product3), their distance exceeds the threshold $D_{min}$. Thus, *ALE* message is sent by node3 to the control center (node0). The message exchange for this case is presented in figure 26.

```
...
node: 3 -> sent GRE on BROADCAST at 38.848126
node: 1 -> received APP_DATA_PACKET(GRE) from node3 at 38.853704
node: 2 -> received APP_DATA_PACKET(GRE) from node3 at 38.853704
node: 4 -> received APP_DATA_PACKET(GRE) from node3 at 38.853704
node: 5 -> received APP_DATA_PACKET(GRE) from node3 at 38.853704
node: 6 -> received APP_DATA_PACKET(GRE) from node3 at 38.853704
node: 6 -> sent RSI to node3 at 38.853911
node: 3 -> received APP_DATA_PACKET(RSI) from node6 at 38.858001
node: 3 -> sent ALE to node0 at 38.858001
node: 0 -> received APP_DATA_PACKET(ALE) from node3 at 38.89755
node: 0 -> sent ACKALE to node3 at 38.89755
node: 3 -> received APP_DATA_PACKET(ACKALE) from node0 at 38.903670
```

Figure 26. Castalia trace file of incompatible distance alert scenario

Upon receiving the *GRE* message of node3 at 38.853704s, node6 verifies the incompatibility security level and sends *RSI* message to node3. Node3 detect danger incompatible state in the instant t1=38.858001s. Thus, *ALE* message is sent to node0 (the sink). Node0 received *ALE* at 38.89755 within a delay of 0.043846s.

## 7. CONCLUSIONS

In this paper, an application and network layers for wireless sensor networks are developed to supervise dangerous chemical products warehouse using the concept of active product. Two critical factors are considered, the ambient temperature of chemical product and their incompatibility. The distance between the incompatible products is supervised to prevent all possible dangerous reaction. Four security rules are proposed to monitor these factors: static, dynamics, community, and global rules. Then, a supervising application layer is developed to exchange messages between the active product and the control center using different communications step: registration, configuration, internal surveillance, and alert announcement. Finally, network layer is developed for routing the different application message. Different routing strategies are proposed for each message class. Node-disjoint multipath routing is used for alert message to provide fault-tolerant and high quality of service. And, information gathering-based routing is used to route the routine messages. Through computer simulation using Castalia/OMNeT++ tools, the performance of our solution is evaluated and studied. Simulation results showed that the network layer can achieve short average end to end delay and very low packet loss rate for alert messages. In addition, the application layer reacted perfectly for critical event (high temperature, incompatible products very close).